\documentclass[
 preprint,
 amsmath,amssymb,
 aps, physrev,
]{revtex4-2}
\usepackage[german, english]{babel}

\usepackage{graphicx}
\usepackage{dcolumn}
\usepackage{bm}
\usepackage[
    colorlinks=true, 
    linkcolor=blue,  
    citecolor=blue,  
    urlcolor=blue    
]{hyperref}

\begin{document}

\title{\textbf{Nonreciprocal spin wave in room-temperature van der Waals ferromagnet $(\rm Fe_{0.78}Co_{0.22})_{5}GeTe_{2}$} 
}%

\author{Guofu Xu$^{1}$} 
\thanks{These authors contributed equally to this work.} 

\author{Feihao Pan$^{2}$} 
\thanks{These authors contributed equally to this work.} 

\author{Jiyang Ou$^{3,4,5}$} 
\thanks{These authors contributed equally to this work.} 

\author{Wenjun Ma$^{1}$}
\author{Xu Zhang$^{1}$}
\author{Xiling Li$^{1}$}
\author{Guoqiang Yu$^{3,4,5}$}

\author{Peng Cheng$^{2}$}
\email{Contact author: pcheng@ruc.edu.cn}

\author{Hongjun Xu$^{6,1,3}$}%
\email{Contact author: xuhongjun@lzu.edu.cn}

\author{Guozhi Chai$^{1}$}%
\email{Contact author: chaigzh@lzu.edu.cn}
\affiliation{$^{1}$ Key Laboratory of Magnetism and Magnetic Functional Materials of the Ministry of Education, Lanzhou University, Lanzhou, 730000, China}

\affiliation{$^{2}$ Laboratory for Neutron Scattering and Key Laboratory of Quantum State Construction and Manipulation (Ministry of Education), School of Physics, Renmin University of China, Beijing 100872, China}

\affiliation{$^{3}$ Beijing National Laboratory for Condensed Matter Physics, Institute of Physics, Chinese Academy of Sciences, Beijing 100190, China}

\affiliation{$^{4}$ Center of Materials Science and Optoelectronics Engineering, University of Chinese Academy of Sciences, Beijing 100049, China}

\affiliation{$^{5}$ Songshan Lake Materials Laboratory, Dongguan 523808, China}

\affiliation{$^{6}$ School of Materials and Energy, Lanzhou University, Lanzhou, 730000, China}

\date{\today}

\begin{abstract}
    
 Here, we investigate the spin waves in room-temperature van der Waals ferromagnet $(\rm Fe_{0.78}Co_{0.22})_{5}GeTe_{2}$ by utilizing Brillouin light scattering technique. The spin wave dispersion in flakes of different thicknesses shows the key role of dipolar interaction in the spin waves of vdW ferromagnets, and the non-reciprocity of spin wave in thick flakes is observed, which is attributed to the bulk Dzyaloshinskii-Moriya interaction after excluding the influence of dynamic dipolar interaction. The measured bulk DMI parameter D is 0.08 $\rm mJ/m^2$, which is double that of pure $\rm Fe_5GeTe_2$. Our  work shows that Co-doped $\rm Fe_5GeTe_2$ is a promising platform for investigating propagating spin wave and topological spin textures at room temperature.

\end{abstract}

\maketitle

\section{Introduction}
  Spin waves\cite{sw1,sw2}, the collective behavior of electron spin excitations, are considered a potential carrier for transmitting, processing, and storing information in the next generation of quantum information and quantum computing\cite{ap1,ap2,ap3,ap4} due to their unique properties, such as the absence of Ohmic dissipation, nonlinearity, nonreciprocity, and chirality\cite{pro1,pro2,pro3}. One of the most important properties is nonreciprocity and chirality, which arises from the dipole-dipole interaction\cite{dd1,dd2}, or Dzyaloshinskii-Moriya interaction\cite{dmi1,dmi2}, which enable unidirectional transmission of spin waves or facilitate the formation of topological spin texture such as skyrmion\cite{ud,sk}respectively. 

  Since the discovery of two-dimensional van der Waals(vdW) materials $\rm CrI_3, Cr2Ge_2Te_6$ \cite{vdw1,vdw2}with long-range magnetic order, unique layered structures and excellent electrical transport, optical, and spin properties, have attracted widespread attention for potential applications in spintronics, magneto-optical devices, and quantum computing\cite{va1,va2}. Compared to classical thin-film systems, magnetic vdW materials offer greater freedom in magnetization manipulation, interface engineering, and investigating the unique physical properties in the low-dimensional limit\cite{vm1,vm2}. However, realizing spintronic devices based on vdW materials and using spin-wave carriers presents a significant obstacles, e.g., the relatively low Curie Temperature\cite{xuh}. Early vdW materials had Curie temperatures far below room temperature, but the discovery of the $\rm Fe_nGeTe_2$ family of vdW materials has overcome this limitation, that the Curie temperature can reach as high as 310 K\cite{tc} in $\rm Fe_5GeTe_2$, which also displays a relatively small magnetic damping factor comparable to the conventional ferromagnetic metals like permalloy\cite{g1}. Furthermore, Co-doped $\rm Fe_5GeTe_2$ have been shown to achieve a Curie temperature as high as 350 K\cite{tc2}. Higher Co doping ratios can induce a phase transition from a ferromagnetic to an antiferromagnetic magnetic order\cite{pt}.

 Although some studies have investigated the spin dynamics of vdw magnetic materials\cite{sd,stack,g1,lin}, they have been limited to ferromagnetic resonance modes with a wave vector $k\sim 0$. Here, we measure the spin waves in room-temperature van der Waals ferromagnet $(\rm Fe_{0.78}Co_{0.22})_{5}GeTe_{2}$ using Brillouin light scattering. The spin wave dispersion in flakes of different thicknesses reveals the key role of magnetic dipolar interaction in vdW spin waves, and the spin wave nonreciprocity phenomenon in thick flakes is observed, which is attributed to the bulk Dzyaloshinskii-Moriya interaction .

\section{Experiment}
  Relatively large-size $(\rm  Fe_{0.78}Co_{0.22})_{5}GeTe_{2}$ (FCGT) flakes with different thickness were obtained by Au assisted mechanically exfoliation\cite{xuh} from a bulk FCGT\cite{tc2}, and were following covered by $\rm MgO/SiO_2$ layers sputtering in high vacuum system to prevent oxidation. We prepared four different FCGT flakes with the dimensions of hundreds micrometers and measured their thickness using atomic force microscopy(AFM) (see Supplementary Fig. S1), two of them are 70 nm and 750 nm, close to the bulk state, and the remaining two are 2 nm and 5 nm, consisting only of the several layers. The in-plane and out-of-plane hysteresis loops were also measured on several of the samples using the magneto-optical Kerr effect (MOKE)(see Supplementary Fig. S2). Brillouin light scattering(BLS) was used to measure spin waves, which has the advantage of simultaneously measuring both the energy and momentum of the spin waves\cite{bls1,bls2}. When a laser beam is focused onto the surface of a sample, the laser photons will be inelastically scattered by quasiparticles (such as magnons, phonons, etc.) in the sample. The backscattering light is collected and sent to a (3+3) channel Fabry–Pérot interferometer(JRS) for frequency analysis, which obtained the energy of the spin waves by comparing backscattering light frequency with the reference laser frequency. The wave vector of the spin wave is determined by the direction of the incident laser, $k_{sw} = 4\pi sin\theta/\lambda_{L}$, where $\theta$ is the angle between the incident laser and the normal direction of the sample plane and $\lambda_{L}$ is the incident laser wavelength(532 nm), which can achieve a maximum value of 20 rad/$\mu$m in our experiments. Moreover, Brillouin light scattering can be used to quantify the strength of the Dzyaloshinskii-Moriya interaction(DMI). When inelastic scattering occurs, there will be a Stokes process of quasi-particle generation and an anti-Stokes process of quasi-particle annihilation. These two processes represent different wave vector directions in the Damon-Eshbach(DE) mode configuration($k_{sw}\perp M$)\cite{de1,de2}, and such chiral interaction will have different influences on frequencies of these two peaks, the DMI strength can be derived from the frequency difference between these two peaks. In our experiments, the applied bias magnetic field was always perpendicular to the spin wave vector, which constitutes the DE mode configuration. The laser power was 45 mW during the experiments and all experiments were measured at room temperature.

\section{Results and discussions}
  The crystal structure of Co-doped $\rm Fe_5GeTe_2$(FGT) is shown in Fig. 1a, which consists of a two-dimensional Fe-Ge slab sandwiched between Te layers, and previous studies\cite{pt,pr1,pr2,stack} have determined  that the space group of FGT is $R\overline{3}$m. In each unit cell, there are three nonequivalent Fe sites, namely Fe1, Fe2 and Fe3, Fe1 occupies a split site, which may be located above or below the Ge atom, and this split site occupancy leads to the uniquely complex crystal structure and magnetic properties of FGT. The Co atoms in Co-doped FGT mainly replace the Fe atoms at the Fe1 site\cite{pr2}. Under low Co concentration doping(0.2 - 0.4), the sample maintains a rhombohedral ABC stacking with a ferromagnetic ground state, but the easy magnetization axis changes from out-of-plane to in-plane. Higher Co doping concentration($>$ 0.45) will cause the sample to undergo a structural phase transition, forming a hexagonal $AA^{\prime}$ stacking and antiferromagnetic order.

  \begin{figure}[htbp!]
\includegraphics[width=0.9\textwidth]{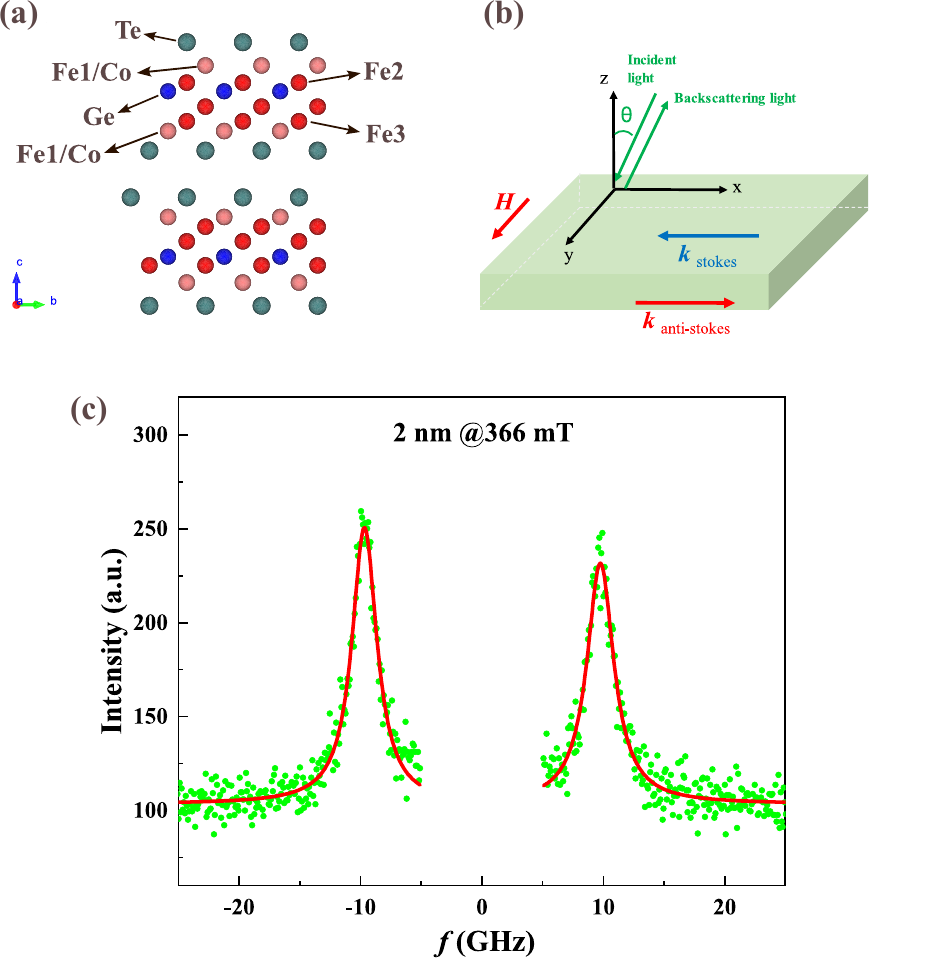}
\caption{\label{fig1} (a) Schematic diagram of the FCGT crystal structure, where the Fe1 site is the splite site, and crystal structure was drawn using the VESTA\cite{cry1} and a cif file from the Computational 2D Materials Database\cite{cry2,cry3}. (b) Schematic diagram of the Brillouin light scattering setup. (c) Brillouin light scattering spectrum of a 2 nm thick FCGT flakes under a magnetic field of 366 mT. The green dots are experimental points, and the red solid line is the result of a Lorentzian fit.}
\end{figure}

  First, we used BLS to measure the thermally excited spin-wave signals of four FCGT flakes under different magnetic fields to determine some static magnetic parameters, such as the gyromagnetic ratio $\gamma$, the saturation magnetization $M_{\rm s}$, the out-of-plane anisotropy $H_{ani}$ et al. Fig. 1b shows the BLS measurement experimental setup, where the direction of the external magnetic field $\boldsymbol{H}$ is parallel to the sample surface plane, and the direction of the in-plane wave vector $\boldsymbol{k_{\parallel}}$ is perpendicular to the direction of the external magnetic field. Fig. 1c shows the Brillouin light scattering spectrum of a 2 nm FCGT flake at the external magnetic field of 366 mT. The Stokes and anti-Stokes peaks are fitted using a Lorentzian fit to obtain parameters such as peak position and linewidth, as shown in Fig. 1c which signifies even for a FCGT flake with a thickness of 2 nm, approximately two layers, nearing the two-dimensional limit, a nice BLS signals can be maintained.

\begin{figure}[htbp!]
\includegraphics[width=0.9\textwidth]{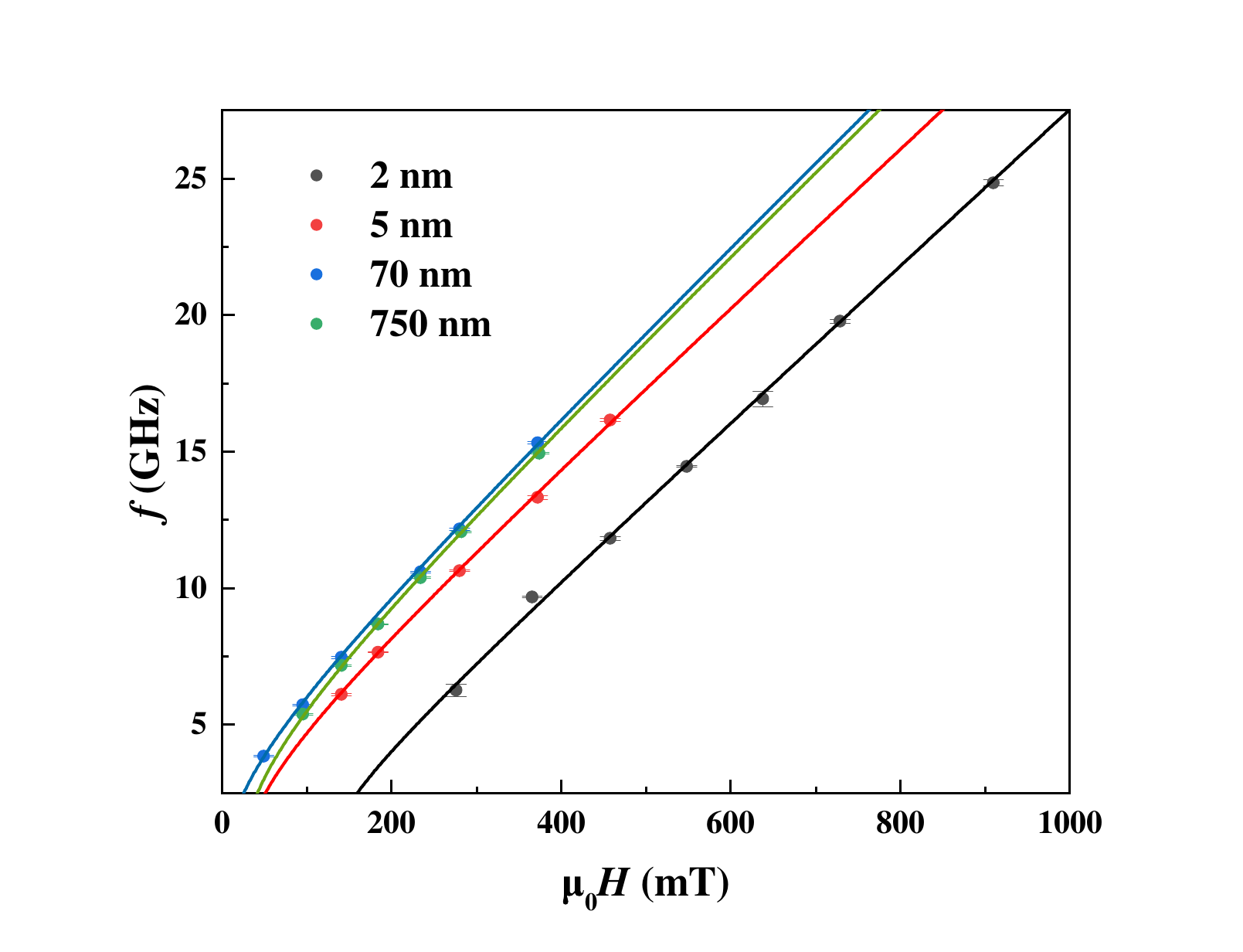}
\caption{\label{fig2} The the spin-wave frequency in FCGT flakes of different thicknesses varies with the magnetic field. The frequency data is obtained by Lorentz fitting the BLS spectrum, and the solid line is obtained by fitting using Eq. (2)}
\end{figure}

  All the data extracted under varying magnetic fields are shown in Fig. 2 , which shows the relationship between the spin wave frequencies under different magnetic fields when $\theta = 20^{\circ} $ (the equivalent wave vector is 8.08 rad/$\mu$m) (Brillouin light scattering spectrums of different samples are shown in Supplementary Material Fig. S3). The static magnetic parameters are fitted and extracted from $f_0$ by using the law of dispersion of magnetostatic spin-waves determined by Kalinikos \& Slavin\cite{ks}:

\begin{equation}
\label{eq1} 
\begin{aligned}
f_0 &= \frac{\gamma\mu_0}{2\pi}
\sqrt{\bigg[H_{\rm eff}+\frac{2A}{\mu_0 M_s}k_{\parallel}^2 + M_s(1-P_0)\bigg]
\bigg[H_{\rm eff}+\frac{2A}{\mu_0 M_s}k_{\parallel}^2 + M_s P_0\bigg]},
\end{aligned}
\end{equation}

\noindent
 where $\gamma$ is the gyromagnetic ratio, $H_{\rm eff}$ is the effective field, including the external magnetic field and the out-of-plane anisotropic field, $ \mu_0$ is the vacuum permeability, $M_s$ is the saturation magnetization, $A$ is the exchange stiffness, $k_{\parallel}$ is the in-plane wave vector, and these two terms represent exchange interaction, which had little effect in our measurements and we ignored them in subsequent analyses. $P_0 = 1-\frac{1 - \exp(-k_{\parallel}t)}{k_{\parallel}t}$ is the geometry factor, which is a very small value under our experimental wave vector,  
 therefore, we simplify Eq. (1) to 
 
\begin{equation}
\label{eq2} 
\begin{aligned}
f_0 &= \frac{\gamma\mu_0}{2\pi}\sqrt{\bigl(H_{\rm ext}-H_{\rm ani}\bigr)\bigl(H_{\rm ext}-H_{\rm ani}+M_{\rm s}\bigr)} \
\end{aligned}
\end{equation}

 \noindent
 to describe and fit the experimental data in Fig. 2, where the magnetic parameters, gyromagnetic ratio $\gamma$, saturation magnetization $M_{\rm s}$, out-of-plane anisotropy $H_{\rm ani}$, extracted by fitting are shown in Table 1.

\begin{table}[htbp!]
\caption{\label{tab1} Magnetic parameters of FCGT flakes with different thicknesses}
\begin{ruledtabular}
\begin{tabular}{
    c 
    D{.}{.}{2.1}
    D{.}{.}{3.0} 
    D{.}{.}{1.3} 
}
\textrm{t (nm)} &
\multicolumn{1}{c}{\textrm{$\frac{\gamma}{2\pi}$ (GHz/T)}} &
\multicolumn{1}{c}{\textrm{$\mu_0 H_{\rm ani}$ (mT)}} &
\multicolumn{1}{c}{\textrm{$\mu_0 M_{\rm s}$ (T)}}\\
\colrule
2 & 28.5 & 126 & 0.193\\
5 & 28.7 & 28 & 0.295\\
70 & 30.5 & 5 & 0.316\\
750 & 30.1 & 24 & 0.360\\
\end{tabular}
\end{ruledtabular}
\end{table}

  The obtained gyromagnetic ratio $\gamma/2\pi$ is in the range of 28.5-30.5 GHz/T, corresponding to a g-factor of 2.04-2.18. Previous studies have reported that in vdW material FGTs, orbital angular momentum contributes to the gyromagnetic ratio due to strong spin-orbit coupling, causing its value to deviate from the g-factor of 2, which is generally understood to be derived from electron spin angular momentum\cite{g1,g2}. Although bulk FCGT is a ferromagnetic state with an easy plane, samples of all thicknesses exhibit a certain out-of-plane anisotropy. This value is not significant in thick films, where the demagnetization field generated by shape anisotropy still dominates, and the magnetic moments tend to align in-plane. However, as the number of layers decreases, FCGT exhibits a strong perpendicular magnetic anisotropy. Besides the demagnetization effect as the thickness decreases and the enhanced perpendicular magnetic anisotropy caused by surface-interface effects similar to those in ferromagnetic ultrathin films, this magnetic anisotropy may be an intrinsic two-dimensional effect near the two-dimensional \cite{ani1,ani2}. Furthermore, in samples with a thickness of 2 nm, approaching the two-dimensional limit, the saturation magnetization decreases rapidly probably due to thermal fluctuations. 

   After obtaining the magnetic parameters of the sample, we then used BLS to measured the spin wave dispersion spectra of samples of different thicknesses. As shown in Fig. 3a, the spin-wave dispersion relations of FCGT flakes at different thicknesses and different magnetic fields are given. First, it can be seen that there are two obvious types of dispersion relations with different trends, one for 70 and 750 nm, and the other for 2 and 5 nm. In the case of thin flakes of 2 and 5 nm, the dispersion shows a nearly flat band feature, while in the thick flake state, it exhibits nearly a linear dispersion relation.

   Here, we consider FCGT, a vdW material with an in-plane easy axis ferromagnetic coupling between layers and within layers, as a ferromagnetic multilayer, and the magnetostatic spin wave mode of the ferromagnetic multilayer has been studied by P. Gr{\"u}nberg and K. Mika\cite{pmika1,pmika2}. For an $N$-layer multilayer film with a magnetic layer thickness of $d$, a magnetization of $M_{\rm s}$, and a non-magnetic layer spacing of $d_{\rm 0}$, the magnetostatic spin wave relation is given by

 \begin{equation}
\label{eq3} 
\begin{aligned}
  f_{0}=\frac{\gamma \mu_{0}}{2 \pi}\left(H^{2}+H \cdot M_{s}+\frac{M_{s}^{2}}{4} \mathcal{W}\right)^{\frac{1}{2}},
\end{aligned}
\end{equation}

 \noindent
 where, $\mathcal{W}$ is a dimensionless geometric factor related to the wave vector $k$, the thickness of the magnetic layer $d$, and the thickness of the spacer layer $d_{\rm 0}$. The solution of $\mathcal{W}$ is determined by an $N\times N$ matrix, which has 2$N$ solutions, indicating that there are 2$N$ magnetostatic spin wave modes in the $N$-layer magnetic multilayer film. These collective modes can be mainly divided into two categories, 2 surface modes and 2$N$-2 bulk modes. When $N$ becomes larger enough, the bulk modes will gradually approach and form an energy band, while the surface modes will no longer change at a certain $N$, which has also been revealed in the recent study of A-type van der Waals antiferromagnetic surface spin waves\cite{ssw}. When $N = 1 $, $\mathcal{W} = 1- {e}^{-2kd}$, which is the form of the classic magnetostatic surface spin wave of an infinite magnetic film with thickness $d$, known as the Damon-Eshbach mode. As BLS measurement is sensitive to surface modes, obvious spin wave signals can be measured in flakes as low as 2 nm. Since there is a clear frequency asymmetry between the positive and negative wave vector in the 70 and 750 nm flakes spin wave dispersion spectra, we add a term to the right side of Eq. (3), S$\times k_{\parallel}$, where S is a constant, which will be discussed in detail below.

\begin{figure}[htbp!]
\includegraphics[width=0.9\textwidth]{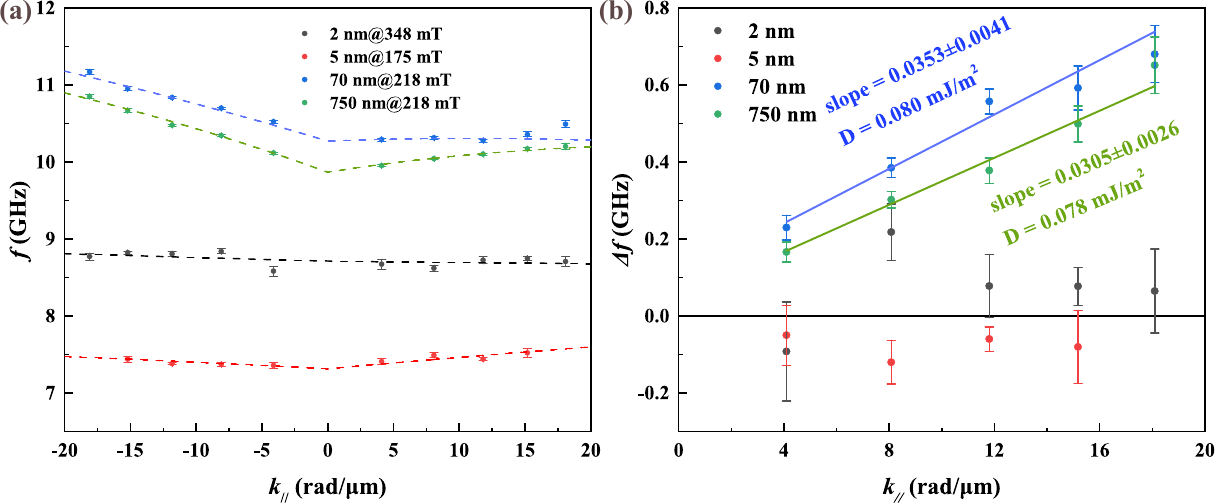}
\caption{\label{fig3} (a)The the spin-wave frequency in FCGT flakes of different thicknesses varies with the in-plane wave vector. The frequency data is obtained by Lorentz fitting the BLS spectrum, and the solid line is obtained by fitting using Eq. (3). (b)The difference between the spin wave frequencies of the Stokes peak and the anti-Stokes peak varies with the in-plane wave vector. The solid line is the result of linear fitting.}
\end{figure}

   Using the static magnetic parameters in Table 1 and Eq. (2), we fitted the dispersion, as shown in Fig. 3a. In 2 and 5 nm flakes, the dipolar interaction is minimal and contributes essentially nothing to the spin wave frequency, resulting in a near-flat band at the long wavelengths of 20 rad/$\mu$m measured. At 70 and 750 nm flakes, the dispersion increases approximately linearly\cite{lin} at small wave vectors due to the dipole contribution. As mentioned above, the surface mode spin waves of 70nm and 750nm thick FCGT flakes have reached a limit and no longer change, the spin wave dispersion no longer varies dramatically with flake thickness as shown in Fig. 3a. Furthermore, another key feature of spin wave dispersion is the nonreciprocity of spin wave frequencies under opposite wave vectors, which is reflected in the Brillouin light scattering spectrum as a frequency asymmetry between Stokes peak and Anti-Stokes peak. Although the measurement itself introduces a systematic error of approximately 0.1 GHz due to the misalignment of the reference and signal beams, the frequency difference is clearly much larger than this value, as shown in Fig. 3b, and this frequency difference is positively correlated with the wave vector. This nonreciprocity mainly has two possible origins. One is the interlayer dipolar interaction, the dynamic dipolar interaction generated by the spin precession of one magnetic layer will affect the spin wave frequency of the other layer\cite{dd1,dd2}. The other is the Dzyaloshinskii-Moriya interaction, the chiral DM interaction has opposite effects on positive and negative wave vectors, leading to  nonreciprocity of spin wave\cite{dmi1,dmi2}. Recently, the observation of skyrmions in FGT and FCGT systems\cite{skyr1,skyr2} suggests the the existence of DMI.

   Many experiments in ferromagnetic bilayers have observed spin wave nonreciprocity caused by dynamic dipolar interaction\cite{nr1,nr2}, which nonreciprocal frequency can reach up to 3-5 GHz, however, it remains unclear whether the dynamic dipolar interaction driven nonreciprocity occurs in stacked vdW materials. Here, we demonstrate through experiment and simulation that  the nonreciprocity induced by interlayer dynamic dipolar interaction does not occur in stacked multilayers with identical saturation magnetizations, a structure similar to vdW materials. We prepared multilayer films of $[\rm CoFeB(7 nm)/SiO_2(4 nm)]_{\it  N}$ by magnetron sputtering, with n ranging from 1 to 4,and the thickness of the spacing layer was thick enough to completely block exchange interactions such as the RKKY interaction, leaving only the dynamic dipolar interaction between layers to affect the spin wave dispersion. The spin wave frequency of the sample was measured by BLS, as shown in Fig. 4a, the difference in spin wave frequency between the opposite wave vector directions is very small and within the measured error range. Furthermore, no linear dependence on the wave vector is observed in the stacks from the monolayer to four-layer. 

   \begin{figure}[htbp!]
\includegraphics[width=0.9\textwidth]{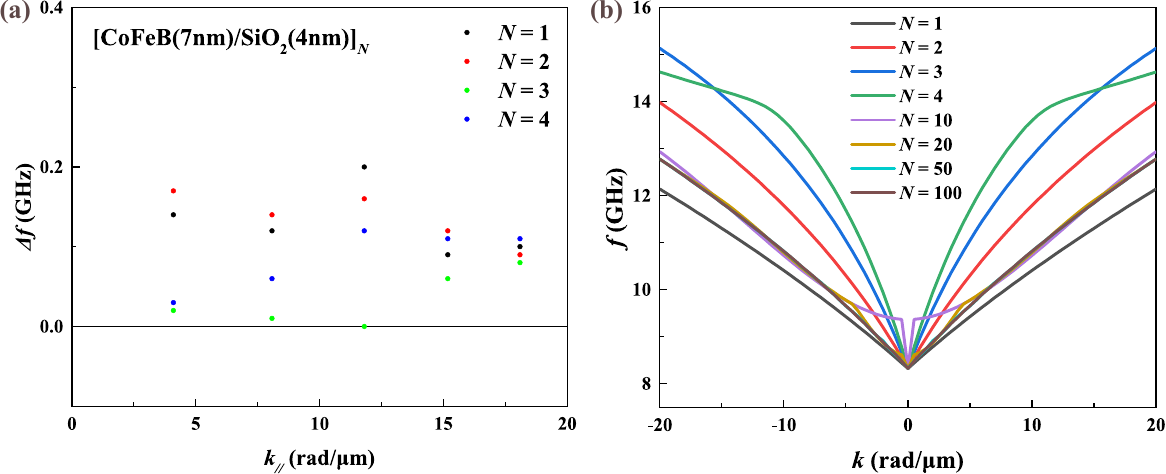}
\caption{\label{fig4} (a) Frequency difference of the $[\rm CoFeB(7 nm)/SiO_2(4 nm)]_{\it  N}$ multilayer films between the spin wave frequencies of the Stokes peak and the anti-Stokes peak varies with the in-plane wave vector. (b)Simulation results of spin wave dispersion in an $N$-layer structure, where each layer has the same thickness of 7 nm and the same saturation magnetization of 950 kA/m, and the spacing between each layer is 4 nm.}
\end{figure}

  Some studies\cite{nr1,nr2} have discussed that the nonreciprocity induced by the dynamic dipolar interactions of multi-layers may be related to the magnetization of different layers. Below we also found from the simulations that a multilayer structure with the same magnetization value in each layer is not sufficient to form nonreciprocity. We designed a $N$-layer multilayer structure similar to the experiments with uniform magnetization, each layer is 7 nm thick, and the interlayer spacing is 4 nm. Spin-wave dispersion was simulated using the open-source finite-element software $TetraX$\cite{tx1,tx2}, which does not rely on the time integration of the Landau-Lifshitz-Gilbert equations, but instead directly solves for the eigenmodes via a propagating-wave dynamic matrix approach. The spin-wave spectrum obtained through simulation is shown in Fig. 4b, and there is no nonreciprocity in this structure. In addition, when $N = 1$ to 4, the simulation results are consistent with our experimentally measured surface mode spin waves (see Supplementary Figure S4). As $N$ gradually increases, the surface spin wave mode gradually converges to a certain mode(See Fig. 4b, the surface mode spin waves of $N = 50$ and $N = 100$ have completely overlapped.), which is consistent with the theoretical model above. What is more important, none of the above dispersion curves shows non-reciprocity, therefore, we determined that the nonreciprocity observed in our experiment in the FCGT vdW system does not come from the interlayer dynamic dipolar interaction, but from the DM interaction.

  The frequency difference $\Delta f = \left|f_{\text {Stokes }}-f_{\text {anti-Stokes }}\right|$ between Stokes peak and anti-Stokes peak can be used to quantify the The frequency difference between s and as can be directly used to quantify the parameter D  of the DMI\cite{dmi1,dmi2}: 
   \begin{equation}
        \Delta f=\frac{2\gamma}{\pi M_s}Dk
    \end{equation}
   \noindent
   where $\gamma$ is the gyromagnetic ratio, $k$ is the wave vector, as shown in Fig. 3b, the value of D is determined by the slope of $\Delta f(k)$. Experimental data indicated a D value of 0.08 $\rm mJ/m^2$ in our measured FCGT system, comparable to many spintronics systems, though still smaller than that of the classical ferromagnets/heavy metal(FM/HM) ultrathin film system. In the FM/HM ultrathin film, the DMI arises from spin-orbit coupling and inversion symmetry breaking at the interface, which is an interfacial DMI, and its intensity significantly depends on the thickness of the magnetic film, exhibiting a $1/t$ decay trend\cite{t1}. However, our experimental results show that the DMI appears in the bulk state and remains essentially unchanged at 70nm and 750nm, while no significant DMI is observed in ultrathin films. Therefore, we attribute the spin-wave nonreciprocity observed in the FCGT to the bulk DMI. The generation of bulk DMI also requires inversion symmetry breaking, but the space group of the FGT-based material is $R\overline{3}$m, which is centrosymmetric. There are two possible mechanisms to explain the bulk DMI of $\rm Fe_nGeTe_2$. One is that Fe vacancies or stacking faults may appear in the FGT\cite{sf1,sf2,sf3}, which will lead to a local symmetry breaking, however, in our experiment, the laser spot size is at the level of hundreds of microns, and the measured DMI is an average result. The second mechanism is symmetry breaking within the FGT vdW layers, each FGT monolayer contains two partially filled Fe sites (split sites), located at different vertical positions: Up(U) and Down(D)\cite{site1,site2,site3}. First-principles calculations have shown that in the “UUU”, "UUD", "UDD" arrangement, such an ordered superstructure can produce bulk DMI\cite{site4,site5}. Recently, the existence of bulk DMI in FGT epitaxial films was experimentally observed, with a measured D value of 0.04 $\rm mJ/m^2$\cite{f}. The D value we measured in Co-doped FGT was twice as large as theirs, further confirming this mechanism, that Co atoms preferentially replaces these Fe splitting sites, which leads to more significant symmetry breaking. A larger DMI strength leads to a smaller skyrmion size, which in turn increases its storage density in spintronic applications.

\section{Conclusion}

In conclusion, we used BLS to measure spin waves in the room-temperature ferromagnet $(\rm Fe_{0.78}Co_{0.22})_{5}GeTe_{2}$. The measurements of FCGT flakes of varying thickness revealed that spin waves exhibit near-flat band characteristics in thinner flakes, while in thicker flakes, the spin waves exhibit near-linear behavior within a small wave vector range due to dipolar interactions. Furthermore, we observed spin-wave nonreciprocity in thicker flakes, and attributed this to the bulk DMI, after eliminating the influence of dynamic dipolar interactions and comparing the spin-wave spectra of FCGTs of varying thicknesses. The bulk DMI observed in FCGT is approximately 0.08 $\rm mJ/m^2$, which may originate from the partially ordered arrangement of Fe split sites. This value is twice as large as that of pure FGT, indicating that doping modulation can effectively adjust the bulk DMI parameter of vdW materials. Our  work shows that Co-doped $\rm Fe_5GeTe_2$ is a promising platform for investigating propagating spin wave and topological spin textures at room temperature.

\section*{Acknowledgements}
This work is supported by the National Natural Science Foundation of China (NSFC) (Nos. 52471200, 12174165, 52201219, 12574124, 12074426, and 12474148) and Scientific Research Innovation Capability Support Project for Young Faculty (ZYGXQNJSKYCXNLZCXM-l19).

\bibliography{reference}

\end{document}